\documentclass[a4paper,fleqn]{cas-sc}

\usepackage[authoryear]{natbib}
\usepackage{amsmath}
\usepackage{graphicx}
\usepackage{amssymb}
\usepackage{bm}
\usepackage{booktabs}
\usepackage{subcaption}
\usepackage{longtable}
\usepackage{lineno}
\usepackage{algorithmicx}
\usepackage{algorithm}
\usepackage{algpseudocode}
\usepackage{xcolor}
\usepackage{tabularx}
\usepackage{etoolbox}

\usepackage{comment}
\usepackage{stfloats}  
\usepackage{enumitem}
\usepackage{float}

\AtBeginEnvironment{tabularx}{\normalsize}

\begin{document}
\let\WriteBookmarks\relax
\shorttitle{Simulation-based multi-objective PID tuning for ASVs}
\shortauthors{W. Li Y. Tan and A. Toffoli}
\title[mode = title]{Simulation-Based Multi-Objective PID Tuning for Autonomous Surface Vessels in Random Sea States}

\author[1]{Weijia Li}[orcid=0009-0008-9342-1619] 
\ead{weijia.li@student.unimelb.edu.au}
\credit{Conceptualization, Methodology, Software, Investigation, Writing - original draft}

\author[2]{Ying Tan}[orcid=0000-0001-8495-0246]   
\ead{yingt@unimelb.edu.au}                        
\credit{Supervision, Methodology, Writing -- review \& editing}

\author[1]{Alessandro Toffoli}[orcid=0000-0003-3112-6856] 
\cormark[1]
\ead{alessandro.toffoli@unimelb.edu.au}           \credit{Supervision, Writing -- review \& editing}

\affiliation[1]{organization={Department of Infrastructure Engineering, The University of Melbourne},  
city={Melbourne},
state={VIC},
postcode={3010},
country={Australia}}

\affiliation[2]{organization={Department of Mechanical Engineering, The University of Melbourne},  
city={Melbourne},
state={VIC},
postcode={3010},
country={Australia}}

\cortext[1]{Corresponding author.}

\begin{abstract}
Autonomous surface vessels commonly use proportional-integral-derivative (PID) controllers for speed and heading, but tuning these controllers to ensure accurate navigation and safety in waves remains challenging. Extremum seeking control (ESC) enables model-free controller tuning by iteratively adjusting controller gains based on measured performance. However, applying it to marine vessels requires extensive sea trials, which are costly and difficult to reproduce. Here we propose an offline tuning framework that applies ESC to a simulation model of the vessel in random waves, balancing tracking accuracy against sideways tilt (heel) during manoeuvres. A calibration procedure separates gain-induced performance changes from wave-induced variability to select the gains to tune and their step sizes. On independent wave realizations, tuning in waves rather than calm water reduces propeller shaft-speed variability by approximately $66\%$, its root-mean-square rate by $40\%$, and speed-tracking cost by $2.5\%$. Multi-objective tuning reveals a trade-off between tracking and heel, and a compromise configuration improves both relative to the initial gains. These findings provide candidate initial gains for subsequent online adaptation on the real vessel.
\end{abstract}

\maketitle
\section{Introduction}
Autonomous surface vessels (ASVs) are increasingly used for maritime transport, where reliable speed and heading control across a wide range of sea conditions is essential for accurate navigation and safe operation~\citep{fossen2011handbook,cc49Wang2019State-of-the-Art,n1xu2026challenges,r8Akdag2024A}. Ocean waves make control especially difficult, as they introduce time-varying disturbances into vessel motion and feedback measurements, which can affect closed-loop control performance and complicate controller design~\citep{fossen2011handbook,Ren2022Data-driven6dof}.

Various control strategies have been developed for marine vessels, including linear, nonlinear, adaptive, robust, and data-driven approaches~\citep{fossen2011handbook,Fossen1999Backstepping,Skjetne2005Adaptive,Do2004RobustAdaptive,Ren2022Data-driven6dof}. Nevertheless, proportional-integral-derivative (PID) control remains widely used due to its simplicity, reliability, and ease of implementation~\citep{fossen2011handbook,zhang2019ship}. Its performance, however, depends critically on appropriate gain selection, which becomes challenging under nonlinear vessel dynamics and environmental disturbances~\citep{astrom2006advanced,killingsworth2006pid,fossen2011handbook}. Wave disturbances also make the measured closed-loop performance vary randomly, which motivates systematic controller tuning under representative sea conditions~\citep{Ren2022Data-driven6dof,An2025}.

Among various automatic optimization methods, extremum seeking control (ESC) has shown potential for model-free PID tuning by iteratively adjusting controller gains based on measured closed-loop performance, without requiring an explicit plant model~\citep{KRSTIC2000sESC,TAN2006ESC,killingsworth2006pid}. By introducing periodic gain perturbations and estimating the gradient of a performance index, ESC progressively adjusts controller parameters towards locally optimal values~\citep{KRSTIC2000sESC,killingsworth2006pid}. However, applying ESC directly to marine vessels requires many sea trials, which are costly, time-consuming, and difficult to reproduce under varying sea conditions~\citep{cc49Wang2019State-of-the-Art,n1xu2026challenges}. Simulation-based controller tuning offers a practical alternative by enabling systematic gain optimization under repeatable operating conditions before physical deployment~\citep{fossen2011handbook,An2025}.

Stochastic wave excitation poses an additional challenge for ESC-based tuning, even in simulation. Vessel motion in waves can be represented by low-frequency manoeuvring dynamics and high-frequency wave-induced responses~\citep{fossen2011handbook,skejic2008unified}. The wave-induced motions introduce fluctuations into the measured feedback signals and consequently the closed-loop performance cost, potentially interfering with ESC gradient estimation and gain adaptation~\citep{fossen2011handbook,stankovic2010extremum}. In particular, performance costs can vary between evaluation periods even when controller gains remain unchanged, obscuring the effects of gain perturbations. This highlights the need to distinguish gain-induced cost changes from background variations when identifying suitable gains for adaptation and determining their update step sizes.

Besides wave-induced variability, controller tuning must also balance tracking performance and operational safety. During large course changes, aggressive steering can improve tracking but increases turning-induced heel through yaw-roll coupling~\citep{ZhangZou2022}. Previous studies have addressed this trade-off through rudder-roll stabilization~\citep{Zhao2019}, roll-constrained path-following control~\citep{ZhangSunLiu2017,Liu2022}, multi-objective control~\citep{Wang2018}, and dedicated heel-reduction actuators~\citep{Liang2023}. These studies show the importance of this trade-off, motivating the inclusion of both tracking performance and turning-induced heel in controller-gain optimization.

To address these challenges, we develop an offline controller-tuning framework that integrates model-free ESC with a two-time-scale vessel-wave simulation model. The model combines low-frequency manoeuvring dynamics with wave-frequency responses obtained from response amplitude operators (RAOs) computed using the boundary element method (BEM), under irregular waves represented by the JONSWAP spectrum~\citep{skejic2008unified,fossen2011handbook,babarit2015,JONSWAPguo2011simulation,yu2026nemoh}. The framework enables systematic tuning of speed PID and heading P controller gains under stochastic wave excitation. At the core of the framework is a segment-wise ESC procedure that updates the controller gains during continuous simulated vessel operation, without resetting the system state. To handle wave-induced fluctuations, a calibration procedure based on the signal-to-noise ratio (SNR) distinguishes gain-induced performance changes from background fluctuations, identifying the gains suitable for adaptation and determining their update step sizes. The framework is then used to investigate the influence of the tuning environment, by comparing controllers tuned in calm water and in stochastic waves against the initial controller under independent wave realizations. Finally, the framework is extended to multi-objective controller tuning through a normalized performance cost that balances tracking performance and turning-induced heel. The effect of the objective weighting on the tuned gains is examined, and a compromise controller is selected using the minimax-regret criterion.

The paper is organized as follows. Section~\ref{sec:system_modeling} presents the two-time-scale vessel-wave model. Section~\ref{sec:control_section} describes the control architecture, the performance cost, and the segment-wise ESC tuning procedure. Section~\ref{sec:simulation} describes the simulation setup and presents the two studies: the comparison between tuning in calm water and in waves, and the trade-off between tracking and turning-induced heel on a zig-zag route. 
Section~\ref{subsec:discussion} discusses the results and limitations.
Section~\ref{sec:conclusion} presents concluding remarks.

\section{Modeling Framework}
\label{sec:system_modeling}

\begin{figure}[pos=htb!]
    \centering
  \includegraphics[width=\linewidth]{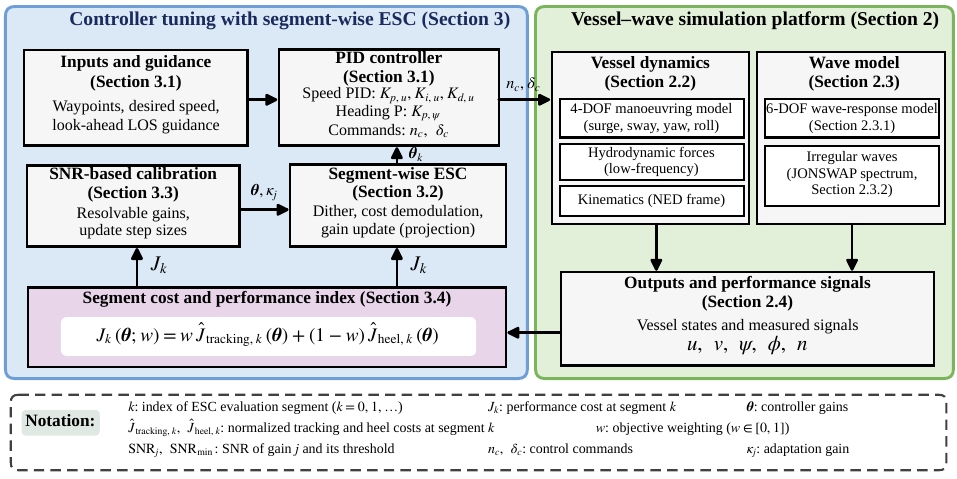}
\caption{Overall structure of the segment-wise extremum-seeking tuning framework, comprising the two-time-scale vessel-wave simulation model, the tracking-heel cost, and the segment-wise ESC loop with its calibration.}
    \label{fig:system_model}
\end{figure}

The proposed offline controller-tuning framework consists of two parts: a vessel-wave simulation model and a controller-tuning loop. The simulation model combines a four-degree-of-freedom manoeuvring model with a six-degree-of-freedom wave-response model under irregular waves, and provides the vessel states and measured signals. In the tuning loop, line-of-sight guidance and the speed PID and heading P controllers generate the propeller shaft-speed and azimuth commands. The resulting performance is evaluated over successive segments through a cost that weights tracking against heel, and the segment-wise ESC updates the controller gains based on this cost, with step sizes set by the SNR-based calibration (Fig.~\ref{fig:system_model}).

\subsection{Modelling Assumptions}
\label{subsec:model_overview}

The two-time-scale modelling framework separates the slowly varying vessel manoeuvring dynamics from the higher-frequency wave-induced motions~\citep{skejic2008unified,fossen2011handbook,perez2005ship}. This separation is based on the following assumptions: (i) the coupling between low-frequency manoeuvring dynamics and high-frequency wave-induced motions is sufficiently weak~\citep{skejic2008unified,perez2005ship}; (ii) first-order wave-induced motions do not directly influence the low-frequency manoeuvring dynamics, while second-order wave-drift loads are neglected~\citep{fossen2011handbook,perez2005ship}; and (iii) ocean waves are considered the primary environmental disturbance, while wind and current effects are excluded to focus on controller performance under stochastic wave excitation. Under these assumptions, the model represents the effects of wave-induced motions on feedback measurements and closed-loop controller performance.

\subsection{Low-Frequency Manoeuvring} Dynamics
The low-frequency module adopts a control-oriented four-degree-of-freedom (4-DOF) manoeuvring model comprising surge, sway, roll, and yaw, following the vectorial formulation of Fossen~\citep{fossen2011handbook}. The roll degree of freedom is retained to capture manoeuvring-induced heel arising from yaw-roll coupling, which is essential for evaluating the trade-off between tracking performance and turning-induced heel considered in this study~\citep{perez2005ship}.

\subsubsection{Low-Frequency Manoeuvring Kinematics}
The low-frequency (LF) manoeuvring dynamics are formulated with respect to an inertial North-East-Down (NED) reference frame and a body-fixed coordinate frame \citep{fossen2011handbook}. Within the proposed two-time-scale framework, low-frequency heave and pitch motions are omitted from the manoeuvring model, since these motions are assumed to be dominated by first-order wave excitation and are represented separately by the high-frequency (HF) wave-response model \citep{fossen2011handbook,perez2005ship}.

Consequently, the vessel’s LF generalized position (pose) is described by
\begin{equation}
\bm{\eta}_{LF}
=
[x,\;y,\;\phi,\;\psi]^T,
\end{equation}
where $x$ and $y$ denote the vessel position in the inertial NED frame, while $\phi$ and $\psi$ represent the roll and yaw angles, respectively. 
The corresponding body-fixed velocity vector is defined as
\begin{equation}
\mathbf{\nu}_{LF}
=
[u,\;v,\;p,\;r]^T,
\end{equation}
where $u$ and $v$ are the surge and sway velocities, and $p$ and $r$ are the roll and yaw rates, respectively.

The corresponding kinematic relationship between the body-fixed velocities and the inertial coordinates is expressed as
\begin{equation}
\dot{\bm{\eta}}_{LF}
=
\bm{J}_{LF}(\phi,\psi)\bm{\nu}_{LF},
\end{equation}
where the reduced 4-DOF kinematic transformation matrix is
\begin{equation}
\bm{J}_{LF}(\phi,\psi)
=
\begin{bmatrix}
\cos(\psi) & -\cos(\phi)\sin(\psi) & 0 & 0\\
\sin(\psi) &  \cos(\phi)\cos(\psi) & 0 & 0\\
0        &  0                & 1 & 0\\
0        &  0                & 0 & 1
\end{bmatrix}.
\end{equation}
Accordingly,
\begin{equation}
\label{eq:kinematics_4dof}
\dot{\bm{\eta}}_{LF}
=
\begin{bmatrix}
u\cos(\psi)-v\cos(\phi)\sin(\psi)\\
u\sin(\psi)+v\cos(\phi)\cos(\psi)\\
p\\
r
\end{bmatrix}.
\end{equation}
The reduced kinematic transformation retains the effect of roll on the horizontal projection of sway velocity through the factor $\cos\phi$. For the moderate roll angles considered, the body-fixed roll and yaw rates are approximated as the corresponding Euler-angle rates, $\dot{\phi}\approx p$ and $\dot{\psi}\approx r$. The resulting formulation represents a control-oriented approximation of the full six-degree-of-freedom vessel kinematics~\citep{fossen2011handbook}.

\subsubsection{Low-Frequency Manoeuvring Kinetics}

The LF manoeuvring kinetics describe the vessel's surge, sway, roll, and yaw responses to propulsion forces, hydrodynamic effects, and hydrostatic restoring moments. Following the vectorial Newton--Euler formulation for marine craft~\citep{fossen2011handbook}, the equations of motion are expressed as
\begin{equation}
\label{eq:kinetics_4dof}
\bm{M}\dot{\bm{\nu}}_{LF}
+
\bm{C}(\bm{\nu}_{LF})\bm{\nu}_{LF}
+
\left[
\bm{D}_{\mathrm{lin}}
+
\bm{D}_{\mathrm{non}}(\bm{\nu}_{LF})
\right]\bm{\nu}_{LF}
+
\bm{g}(\bm{\eta}_{LF})
=
\bm{\tau}_{\mathrm{prop}}
+
\bm{\tau}_{\mathrm{coup}} .
\end{equation}
Here, $\bm{M}\in\mathbb{R}^{4\times4}$ denotes the generalized mass matrix, incorporating rigid-body inertia and hydrodynamic added mass, while $\bm{C}(\bm{\nu}_{LF})\in\mathbb{R}^{4\times4}$ represents the Coriolis and centripetal effects~\citep{fossen2011handbook}. The matrices $\bm{D}_{\mathrm{lin}}\in\mathbb{R}^{4\times4}$ and $\bm{D}_{\mathrm{non}}(\bm{\nu}_{LF})\in\mathbb{R}^{4\times4}$ represent the linear and nonlinear hydrodynamic damping, respectively~\citep{fossen2011handbook,faltinsen1993sea}. The vector $\bm{g}(\bm{\eta}_{LF})$ accounts for hydrostatic restoring forces and moments, primarily associated with vessel roll in the present 4-DOF formulation.

Equation~\eqref{eq:kinetics_4dof} determines the body-fixed accelerations $\dot{\bm{\nu}}_{LF}$. Integration yields the corresponding velocities $\bm{\nu}_{LF}$, which are subsequently transformed into the vessel's LF pose $\bm{\eta}_{LF}$ through the kinematic relationship in Eq.~\eqref{eq:kinematics_4dof}.

The hydrodynamic damping is represented by linear and nonlinear components~\citep{fossen2011handbook,faltinsen1993sea}. The linear component $\bm{D}_{\mathrm{lin}}$ accounts for approximately linear hydrodynamic resistance, while the dominant quadratic damping effects are represented using the diagonal approximation
\begin{equation}
\label{eq:nonlinear_damping}
\bm{D}_{\mathrm{non}}(\bm{\nu}_{LF})
=
\mathrm{diag}
\left(
X_{u|u|}|u|,
Y_{v|v|}|v|,
K_{p|p|}|p|,
N_{r|r|}|r|
\right),
\end{equation}
where $X_{u|u|}$, $Y_{v|v|}$, $K_{p|p|}$, and $N_{r|r|}$ denote the quadratic damping coefficients associated with surge, sway, roll, and yaw, respectively. Each nonlinear damping contribution is therefore proportional to $|\nu_i|\nu_i$, representing velocity-dependent hydrodynamic resistance.

The hydrostatic restoring vector is dominated by the restoring moment associated with vessel roll and is expressed as
\begin{equation}
\label{eq:restoring_4dof}
\bm{g}(\bm{\eta}_{LF})
=
\begin{bmatrix}
0\\
0\\
\Delta g\,\overline{GM}\sin(\phi)\\
0
\end{bmatrix},
\end{equation}
where $\Delta$ denotes the vessel displacement mass, $g$ is gravitational acceleration, $\overline{GM}$ is the transverse metacentric height, and $\phi$ is the roll angle.

The generalized propulsion forces and moments are represented by
\begin{equation}
\label{eq:propulsion_force}
\bm{\tau}_{\mathrm{prop}}
=
\begin{bmatrix}
F_x\\
F_y\\
0\\
N
\end{bmatrix},
\end{equation}
where $F_x$ and $F_y$ are the surge and sway forces generated by the azimuthing propulsion system, and $N$ is the corresponding yaw moment. The zero roll-moment component reflects the assumption that the propulsion system does not directly generate a roll moment in the adopted model. These forces and moments are determined by the propeller shaft speed $n$ and azimuth angle $\delta$, which are regulated by the speed and heading controllers, respectively. The propulsion model and actuator constraints are detailed in Section~\ref{subsec:thruster}.

The coupling between lateral manoeuvring and vessel roll is represented by
\begin{equation}
\label{eq:roll_coupling}
\bm{\tau}_{\mathrm{coup}}
=
\begin{bmatrix}
0\\
0\\
K_v v + K_r r\\
0
\end{bmatrix},
\end{equation}
where $K_v$ and $K_r$ are the hydrodynamic coupling coefficients associated with sway velocity $v$ and yaw rate $r$, respectively. This term represents the roll moment induced by lateral manoeuvring and captures the influence of sway and yaw dynamics on turning-induced heel~\citep{perez2005ship}. 

\subsection{High-Frequency Wave-Induced Motions}
Within the two-time-scale framework, the HF model represents first-order wave-induced vessel motions in six degrees of freedom, complementing the LF manoeuvring model~\citep{fossen2011handbook,skejic2008unified}. The HF response is modelled using linear seakeeping theory and characterized by complex response amplitude operators (RAOs), which describe the amplitude and phase of vessel motion relative to incident waves~\citep{fossen2011handbook,perez2005ship}. The modelling procedure consists of two stages: (i) offline computation of the hydrodynamic characteristics and RAOs over prescribed wave frequencies, vessel speeds, and relative wave headings; and (ii) time-domain reconstruction of irregular wave-induced motions through spectral superposition.

\subsubsection{Hydrodynamic Response Database}
A frequency-domain hydrodynamic analysis is performed offline with the open-source BEM solver NEMOH~\citep{babarit2015} to construct a database of RAOs over the prescribed operating conditions. For each degree of freedom $k\in\{1,\ldots,6\}$, corresponding to surge, sway, heave, roll, pitch, and yaw, respectively, the RAO is expressed as
\begin{equation}
H_k = H_k(\omega,U,\chi),
\end{equation}
where $\omega$ denotes the wave angular frequency, $U$ is the vessel forward speed, and $\chi$ is the relative wave heading.

During time-domain simulation, the instantaneous LF manoeuvring state determines the vessel speed $U(t)$ and relative wave heading $\chi(t)$. The corresponding RAOs are obtained by interpolation from the precomputed database, allowing the HF wave response to vary with the vessel’s operating conditions. This approach accounts for the influence of LF manoeuvring on wave-induced motions without introducing first-order wave excitation directly into the LF equations of motion.

\subsubsection{Time-Domain Wave Reconstruction}
\label{subsubsec:wave_reconstruction}
The irregular sea state is represented by the JONSWAP spectrum $S(\omega)$, which characterizes the distribution of wave energy over frequency~\citep{hasselmann1973measurements}. Only unidirectional (long-crested) waves are considered. The continuous wave spectrum is discretized into $N$ harmonic components. For the $j$th component with angular frequency $\omega_j$, the wave amplitude is given by
\begin{equation}
A_j
=
\sqrt{2S(\omega_j)\Delta\omega},
\end{equation}
where $\Delta\omega$ denotes the frequency interval. Each component is assigned an independent random phase uniformly distributed over $[0,2\pi]$, $\epsilon_j\sim\mathcal{U}[0,2\pi]$.

Following the linear superposition principle~\citep{faltinsen1993sea}, the HF wave-induced motion in the $k$th degree of freedom is reconstructed by superposing the harmonic responses:
\begin{equation}
\label{eq:wave_reconstruction}
\eta_{HF,k}(t)
=
\sum_{j=1}^{N}
A_j
\left|
H_k(\omega_j,U,\chi)
\right|
\cos
\left[
\omega_j t
+
\epsilon_j
+
\angle H_k(\omega_j,U,\chi)
\right].
\end{equation}

For analytical differentiation, the response associated with each harmonic component is expressed in complex form as
\begin{equation}
\mathcal{X}_{j,k}(t)
=
A_j H_k(\omega_j,U,\chi)
e^{i(\omega_j t+\epsilon_j)},
\end{equation}
such that
\begin{equation}
\eta_{HF,k}(t)
=
\mathrm{Re}
\left\{
\sum_{j=1}^{N}
\mathcal{X}_{j,k}(t)
\right\}.
\end{equation}

Since the vessel speed $U(t)$ and relative wave heading $\chi(t)$ vary on the slower manoeuvring time scale, the interpolated RAOs are assumed locally constant over each wave period when differentiating the harmonic responses. Under this approximation, the wave-induced velocities and accelerations are obtained analytically as
\begin{equation}
\label{eq:wave_velocity}
\dot{\eta}_{HF,k}(t)
=
\mathrm{Re}
\left\{
\sum_{j=1}^{N}
i\omega_j\mathcal{X}_{j,k}(t)
\right\},
\end{equation}
and
\begin{equation}
\label{eq:wave_acceleration}
\ddot{\eta}_{HF,k}(t)
=
\mathrm{Re}
\left\{
\sum_{j=1}^{N}
-\omega_j^2\mathcal{X}_{j,k}(t)
\right\}.
\end{equation}
This analytical formulation avoids numerical differentiation of the reconstructed wave motions and provides estimates of the wave-induced velocities and accelerations for subsequent closed-loop simulation and performance evaluation.

\subsection{Integrated Vessel Model}
\label{subsec:integrated_model}

The total vessel motion is obtained by superimposing the LF manoeuvring and HF wave-induced responses, with the 4-DOF LF pose vector first extended to six degrees of freedom as
\begin{equation}
\bm{\eta}_{LF}^{(6)}
=
\begin{bmatrix}
x_{LF} &
y_{LF} &
0 &
\phi_{LF} &
0 &
\psi_{LF}
\end{bmatrix}^{T}.
\end{equation}
The HF wave-induced pose obtained from spectral reconstruction is
\begin{equation}
\bm{\eta}_{HF}
=
\begin{bmatrix}
x_{HF} &
y_{HF} &
z_{HF} &
\phi_{HF} &
\theta_{HF} &
\psi_{HF}
\end{bmatrix}^{T}.
\end{equation}
The total six-degree-of-freedom vessel pose is reconstructed through linear superposition:
\begin{equation}
\label{eq:total_motion}
\bm{\eta}(t)
=
\bm{\eta}_{LF}^{(6)}(t)
+
\bm{\eta}_{HF}(t),
\end{equation}
where
\begin{equation}
\bm{\eta}
=
\begin{bmatrix}
x & y & z & \phi & \theta & \psi
\end{bmatrix}^{T}.
\end{equation}
Accordingly, the horizontal position, roll, and yaw contain both LF manoeuvring and HF wave-induced components, whereas heave and pitch are represented entirely by the HF model.

The HF motion rates and accelerations are obtained analytically from Eqs.~\eqref{eq:wave_velocity} and \eqref{eq:wave_acceleration}. Together with the reconstructed vessel pose and the body-fixed LF velocity vector $\bm{\nu}_{LF}=[u,v,p,r]^T$, these quantities form the motion signals used for feedback control and performance evaluation. 

To represent measurement uncertainty, zero-mean Gaussian noise is added to the feedback signals before they are supplied to the guidance and control algorithms. The prescribed standard deviations are $\sigma_{\eta}=0.05$~m for position, $\sigma_{\phi}=\sigma_{\theta}=0.0009$~rad for roll and pitch, $\sigma_{\psi}=0.0017$~rad for heading, and $\sigma_{\nu}=0.02$~m/s for velocity measurements.

\subsection{Azimuth Thruster Model and Underactuation} 
\label{subsec:thruster}

%The vessel is equipped with a single azimuthing thruster that generates propulsion and steering forces through thrust vectoring. 
The vessel is equipped with a single equivalent azimuthing thruster that generates propulsion and steering forces through thrust vectoring. 

The propeller thrust is calculated using the standard open-water propeller relationship~\citep{fossen2011handbook}:
\begin{equation}
T=\rho_w n^2 D_p^4 K_T(J_a),
\end{equation}
where $T$ denotes the propeller thrust, $\rho_w$ is seawater density, $n$ is the propeller shaft speed (rev/s), $D_p$ is the propeller diameter, and $K_T$ is the thrust coefficient. The advance ratio is defined as
\begin{equation}
J_a=\frac{u}{nD_p},
\end{equation}
where $u$ is the surge velocity. The thrust coefficient $K_T(J_a)$ is approximated by a linear polynomial based on the propeller open-water characteristics~\citep{fossen2011handbook}.
The shaft speed is limited to $[0,150]$~rpm and its rate of change to $1.25$~rpm/s, representing the operating limitations and finite response of the propulsion system.

The azimuthing mechanism directs the propeller thrust through the azimuth angle $\delta$, producing surge and sway forces given by
\begin{equation}
F_x=T\cos\delta,
\qquad
F_y=T\sin\delta,
\end{equation}
and the corresponding yaw moment
\begin{equation}
N=l_tF_y,
\end{equation}
where $l_t$ is the moment arm between the thruster and the vessel centre of gravity. The steering angle is constrained by
\begin{equation}
|\delta|\le35^\circ,
\end{equation}
representing the prescribed mechanical steering limit.

Although the LF manoeuvring model includes four degrees of freedom (surge, sway, roll, and yaw), the propulsion system has only two independent control inputs: propeller shaft speed $n$ and azimuth angle $\delta$, so the vessel is underactuated. The resulting surge and sway forces and yaw moment enter the vessel dynamics through $\bm{\tau}_{\mathrm{prop}}$. Since the sway force $F_y$ and yaw moment $N$ originate from the same thrust vector, they are coupled through the fixed thruster moment arm and cannot be independently commanded. Roll is not independently actuated either, and is influenced indirectly through the manoeuvring-roll coupling $\bm{\tau}_{\mathrm{coup}}$ defined in Eq.~\eqref{eq:roll_coupling}. The actuation and hydrodynamic coupling structure is summarized in Table~\ref{tab:actuation_structure}.
\begin{table}[t]
\centering
\caption{Actuation and coupling structure of the low-frequency manoeuvring model.}
\label{tab:actuation_structure}
\begin{tabular}{ccc}
\hline
Degree of freedom & Propulsion input & Hydrodynamic coupling \\
\hline
Surge & $F_x$ & $0$ \\
Sway & $F_y$ & $0$ \\
Roll & $0$ & $K_v v+K_r r$ \\
Yaw & $N$ & $0$ \\
\hline
\end{tabular}
\end{table}

\section{Controller Tuning with Segment-Wise ESC}
\label{sec:control_section}

\subsection{Control Architecture}
\label{subsec:pid_control}

To investigate controller-gain tuning under stochastic wave excitation, a conventional feedback control architecture is adopted, consisting of a PID controller for speed regulation and a proportional (P) controller for heading control. This relatively simple structure allows the effects of controller-gain adjustments to be examined without introducing the additional complexity of advanced control strategies. The speed controller generates the commanded propeller shaft speed $n_c$, while the heading controller generates the commanded azimuth angle $\delta_c$. Both commands are subject to the actuator constraints described in Section~\ref{subsec:thruster} before being applied to the azimuthing propulsion system.

The desired vessel speed $U_d$ is specified as a constant setpoint, while the desired heading $\psi_d$ is generated using a waypoint-based line-of-sight (LOS) guidance law \citep{moreira2007path,breivik2005guidance}. The LOS look-ahead distance influences the responsiveness of the heading reference and, consequently, the vessel's tracking and manoeuvring behaviour. Since the objective of this study is controller-gain optimization rather than guidance-parameter optimization, the LOS parameters are held fixed throughout the tuning and evaluation procedures. This provides a consistent guidance configuration for comparing the effects of different controller gains.

The reference path is defined by an ordered sequence of waypoints $\bm{p}_m=[x_m\;\;y_m]^T$, $m=1,\ldots,N_{wp}$. For the active segment $\bm{p}_m\rightarrow\bm{p}_{m+1}$, the look-ahead point is defined as
\begin{equation}
\bm{\ell}(t)
=
[\ell_x(t)\;\;\ell_y(t)]^T
=
\bm{h}(t)
+
D_{ah}
\frac{\bm{p}_{m+1}-\bm{p}_m}
{\lVert\bm{p}_{m+1}-\bm{p}_m\rVert},
\end{equation}
where $\bm{h}(t)$ is the orthogonal projection of the measured vessel position $\bm{p}(t)=[x(t)\;\;y(t)]^T$ onto the line containing the active path segment, and $D_{ah}$ is the fixed look-ahead distance. The desired heading is calculated as the bearing from the vessel to the look-ahead point:
\begin{equation}
\label{eq:psi_ref}
\psi_d(t)
=
\operatorname{atan2}
\left(
\ell_y(t)-y(t),
\ell_x(t)-x(t)
\right).
\end{equation}
The active segment advances when the projection point enters the acceptance radius $R_{\mathrm{acc}}$ of the next waypoint, i.e., $\lVert\bm{p}_{m+1}-\bm{h}(t)\rVert<R_{\mathrm{acc}}$. This projection-based switching criterion determines waypoint transitions from along-path progress independently of cross-track error. The simulation terminates when the final waypoint is reached. The guidance and waypoint-switching logic follows the implementation of \citet{le2022shipsimulation}, adapted to the present vessel model.

The speed and heading controllers use the corresponding tracking errors to generate the actuator commands. For speed regulation, the tracking error is defined as
\begin{equation}
e_u(t)=U_d-U(t),
%\qquad
%U(t)=\sqrt{u^2(t)+v^2(t)},
\end{equation}
where 
\begin{equation}
U(t)=\sqrt{u^2(t)+v^2(t)}
\end{equation}
is the magnitude of the measured horizontal vessel velocity. The commanded propeller shaft speed is generated by the PID controller:
\begin{equation}
\label{eq:speed_pid}
n_c(t)
=
K_{p,u}e_u(t)
+
K_{i,u}\int_0^t e_u(\tau)\,d\tau
+
K_{d,u}\dot{e}_u(t),
\end{equation}
where $K_{p,u}$, $K_{i,u}$, and $K_{d,u}$ are the proportional, integral, and derivative gains, respectively. The command $n_c$ is expressed in rpm and converted to rev/s in the propulsion model. The integral contribution is limited to $\pm150$ rpm to mitigate integrator wind-up, and negative shaft-speed commands are set to zero.

For heading regulation, the tracking error is defined as
\begin{equation}
e_\psi(t)
=
\operatorname{wrap}\left(\psi_d(t)-\psi(t)\right),
\end{equation}
where $\operatorname{wrap}(\cdot)$ maps the heading error to the principal angular interval. The commanded azimuth angle is generated by the proportional controller:
\begin{equation}
\label{eq:heading_law}
\delta_c(t)=K_{p,\psi}e_\psi(t),
\end{equation}
where $K_{p,\psi}$ is the heading proportional gain. A proportional controller is adopted to maintain a simple heading-control structure while avoiding the additional sensitivity to wave-induced heading fluctuations and measurement noise associated with derivative action. Integral action is not included because persistent environmental yaw disturbances due to wind, current, and wave drift are excluded from the present model. Any remaining heading-tracking error is accounted for in the performance cost used for controller tuning.

The control architecture comprises four candidate controller gains, collected in the parameter vector
\begin{equation}
\label{eq:theta_all}
\bm{\theta}_{\mathrm{all}}
=
\begin{bmatrix}
K_{p,u} & K_{i,u} & K_{d,u} & K_{p,\psi}
\end{bmatrix}^{T}.
\end{equation}
Although the control architecture contains four controller gains, adapting all of them is not necessarily beneficial. Different gains may have different levels of influence on the measured closed-loop performance, and under stochastic wave excitation, the effects of some gain adjustments may be difficult to distinguish from background fluctuations. An SNR-based calibration procedure (Section~\ref{subsec:snr_calibration}) therefore selects for ESC adaptation only the gains whose effects are sufficiently detectable, while the remaining gains are held at their initial values.

\subsection{Segment-Wise Extremum-Seeking Tuning under Stochastic Waves}
\label{subsec:esc_tuning}

Following the identification of controller gains suitable for adaptation, ESC is employed to adjust their values based on measured closed-loop performance. In the discrete-time ESC approach of \citet{killingsworth2006pid}, controller gains are periodically perturbed, and the resulting variations in measured performance are used to estimate a descent direction for iterative gain adjustment. Although simulation allows repeated evaluations under prescribed conditions, the objective is to investigate a tuning procedure that adjusts controller gains during continuous vessel operation without requiring repeated experiments from identical initial conditions.

Variations in the reference signals may also introduce changes in the measured performance cost that are unrelated to controller-gain perturbations. To improve the comparability of successive performance evaluations, reference signals with well-defined long-term time averages are considered. Specifically, a reference signal $s(t)$ is assumed to satisfy
\begin{equation}
\label{eq:reference_average}
\bar{s}
=
\lim_{T\rightarrow\infty}
\frac{1}{T}
\int_0^T s(t)\,dt
\in\mathbb{R}.
\end{equation}

This class includes constant and periodic references, as well as periodic references with exponentially decaying transient components. Constant speed and heading references, together with prescribed waypoint-following and repeated manoeuvring trajectories, are commonly considered in marine vessel guidance and control \citep{moreira2007path,OhSun2010}.

Based on these considerations, a segment-wise ESC procedure is adopted, in which continuous simulated vessel operation is divided into consecutive evaluation segments. The controller gains remain fixed within each segment and are updated at its end using the measured segment performance. Each segment therefore serves as one closed-loop performance evaluation for the discrete-time ESC algorithm without resetting the vessel state between successive gain updates.

To improve the reliability of segment-wise ESC tuning under stochastic wave excitation, three complementary mechanisms are employed: a settling interval within each segment, high-pass filtering of the segment costs, and an SNR-based calibration performed before ESC training.

Each segment begins with an initial settling interval to reduce the influence of closed-loop transients following controller-gain changes. Specifically, segment $k=0,1,\ldots$ occupies the interval $[t_k,t_k+T_{\mathrm{seg}})$, where $t_k=t_0+kT_{\mathrm{seg}}$ and $T_{\mathrm{seg}}$ is the segment duration. Performance is evaluated over the segment’s final $T_s$ seconds, defined by
\begin{equation}
\label{eq:window_0}
\mathcal{W}_k
=
[t_k+T_{\mathrm{seg}}-T_s,\,
t_k+T_{\mathrm{seg}}).
\end{equation}
The initial interval of duration $T_{\mathrm{seg}}-T_s$ allows transient effects following a gain update to diminish before performance evaluation, assuming local closed-loop stability. The scoring-window duration is chosen as $T_s=N_wT_w$, where $T_w=2\pi/\Delta\omega$ denotes the nominal recurrence time associated with the discretized wave spectrum. This allows performance to be evaluated over multiple nominal wave recurrence periods, reducing sensitivity to short-term wave-phase fluctuations under the adopted harmonic approximation.

The sequence of segment performance measurements $J_k$ is then processed using a discrete-time high-pass filter,
\begin{equation}
\label{eq:esc_filter}
H(z)
=
\frac{z-1}{z+\beta},
\qquad 0<\beta<1.
\end{equation}
The filter removes the constant component and attenuates slowly varying fluctuations in the measured performance cost, while retaining variations at the ESC perturbation frequencies. The parameter $\beta$ controls the frequency response of the filter and is selected to preserve sufficient sensitivity to gain-induced cost variations. The corresponding filtered cost is computed recursively as
\begin{equation}
\label{eq:esc_filter_recursion}
\tilde{J}_k
=
J_k-J_{k-1}-\beta\tilde{J}_{k-1}.
\end{equation}
The filtered sequence $\tilde{J}_k$ is subsequently correlated with the periodic gain perturbations to estimate the local performance gradient \citep{killingsworth2006pid,ariyur2003realtime}. Background fluctuations that remain after settling and filtering are addressed by the SNR-based calibration described in Section~\ref{subsec:snr_calibration}.

With these mechanisms established, the controller gains are updated using the discrete-time ESC formulation of \citet{killingsworth2006pid}. 
For each adapted controller gain $\theta_j$, an admissible interval $[\theta_j^{\min},\theta_j^{\max}]$ is prescribed to constrain the ESC search. In engineering practice, such intervals are often selected based on prior experience and preliminary trial-and-error tuning, taking into account closed-loop performance and actuator limitations. The intervals should be sufficiently broad to allow meaningful gain adjustments while excluding controller settings that produce unacceptable closed-loop responses. The width of each interval is defined as
\begin{equation}
\label{eq:gain_interval_width}
s_j=\theta_j^{\max}-\theta_j^{\min}.
\end{equation}
The gain applied during segment $k$ is expressed as
\begin{equation}
\label{eq:esc_update}
\theta_{j,k}
=
\theta_{j,0}
+
s_j\left[
c_{j,k}+a_j\cos(\Omega_j k)
\right],
\end{equation}
where $\theta_{j,0}$ is the initial gain, $c_{j,k}$ is the dimensionless nominal adjustment ($c_{j,0}=0$), and $a_j$ and $\Omega_j$ denote the normalized perturbation amplitude and frequency, respectively. The scaling factor $s_j$ converts the dimensionless adjustment and perturbation into the corresponding physical gain values.

At the end of each segment, the nominal adjustment is updated according to
\begin{equation}
\label{eq:esc_nominal_update}
c_{j,k+1}
=
\Pi_j\left[
c_{j,k}
-
\kappa_j a_j\cos(\Omega_j k)\tilde{J}_k
\right],
\end{equation}
where $\kappa_j$ is the adaptation gain determined by calibration and $\Pi_j$ is the projection operator enforcing the admissible gain limits.

To ensure that the applied gain remains within its admissible interval, including the periodic perturbation, define
\begin{equation}
c_j^{\min}=
\frac{\theta_j^{\min}-\theta_{j,0}}{s_j},
\qquad
c_j^{\max}=
\frac{\theta_j^{\max}-\theta_{j,0}}{s_j}.
\end{equation}
A projection that reserves sufficient margin for the perturbation is
\begin{equation}
\label{eq:esc_projection}
\Pi_j(c)
=
\min\left\{
c_j^{\max}-a_j,\,
\max\left(c_j^{\min}+a_j,c\right)
\right\}.
\end{equation}
For $0\leq a_j\leq 1/2$, this projection ensures that the applied gain remains within $[\theta_j^{\min},\theta_j^{\max}]$, provided the initial nominal adjustment is admissible.

The perturbation frequencies are selected as $\Omega_j=2\pi b_j/N_d$, where $N_d$ denotes the number of segments in a complete dither period and the integers $b_j$ are chosen to provide distinct, non-aliased perturbation frequencies for different gains \citep{ariyur2003realtime}.

To avoid discontinuities in the commanded propeller shaft speed following gain changes at segment boundaries, the integral state of the speed controller is adjusted to ensure a bumpless parameter update \citep{astrom2006advanced}. The tuning procedure is repeated independently over $N_r$ prescribed training wave realizations, and the final controller gains are averaged across these runs.

\subsection{SNR-Based Calibration for Gain Selection and Adaptation}
\label{subsec:snr_calibration}

The SNR-based calibration procedure is performed before ESC training to identify which of the four candidate controller gains defined in Eq.~\eqref{eq:theta_all} are suitable for adaptation and to determine their corresponding adaptation gains $\kappa_j$ in Eq.~\eqref{eq:esc_nominal_update}. For each candidate gain, the procedure compares the performance variation induced by a prescribed gain perturbation with the background variation observed under fixed controller gains. Gains whose effects cannot be sufficiently distinguished from stochastic fluctuations are excluded from adaptation, thereby reducing the dimension of the tuning problem.

To estimate the gain-induced performance variation, each candidate gain $\theta_j$ is perturbed by $\pm s_ja_j$ in two separate simulations, with the gain set to $\theta_{j,0}+s_ja_j$ and $\theta_{j,0}-s_ja_j$, respectively, while the remaining controller gains are held at their initial values. Here, $s_j$ is the admissible interval width defined in Eq.~\eqref{eq:gain_interval_width}, and $a_j$ is the normalized perturbation amplitude introduced in Eq.~\eqref{eq:esc_update}. The perturbation magnitude $s_ja_j$ is prescribed within the admissible gain limits to produce a measurable performance change while maintaining acceptable closed-loop behaviour.

Let $J_{j,r}^{+}$ and $J_{j,r}^{-}$ denote the corresponding performance costs evaluated over the scoring window defined in Eq.~\eqref{eq:window_0}. The index $r=1,\ldots,N_r$ denotes the training wave realization, with each realization generated using an independent set of random phases for the harmonic wave components while maintaining the same JONSWAP spectral parameters. For each realization, the paired simulations start from the same initial vessel state and use identical wave conditions and scoring windows. This matched setup reduces the influence of environmental differences on the estimated effect of the gain perturbation.

The gain-induced performance variation is estimated as
\begin{equation}
\label{eq:calibration_signal}
d_j
=
\frac{1}{2N_r}
\sum_{r=1}^{N_r}
\left(
J_{j,r}^{+}-J_{j,r}^{-}
\right).
\end{equation}
The quantity $d_j$ represents the average half-difference in performance between the positive and negative perturbations, providing a finite-difference measure of the local influence of gain $\theta_j$ on closed-loop performance.

To quantify the background performance variation, the controller gains are held at their initial values for an additional $N_b$ simulation segments. The resulting segment-cost sequence is processed using the high-pass filter defined in Eq.~\eqref{eq:esc_filter}. Let $\tilde{J}_k^{\mathrm{bg}}$ denote the filtered cost obtained during these fixed-gain segments. The background variation is characterized by its root-mean-square value,
\begin{equation}
\label{eq:esc_background}
\sigma_{\tilde{J}}
=
\sqrt{
\frac{1}{N_b}
\sum_{k=1}^{N_b}
\left(\tilde{J}_k^{\mathrm{bg}}\right)^2
}.
\end{equation}
This quantity measures the magnitude of the filtered performance fluctuations that remain even when the controller gains are unchanged.

The detectability of each gain-induced performance variation is assessed by comparing its estimated magnitude $|d_j|$ with the background fluctuations measured under fixed controller gains. Following the filtering and demodulation mechanism introduced in Section~\ref{subsec:esc_tuning}, the predicted SNR over one dither period is defined as
\begin{equation}
\label{eq:esc_snr}
\mu_j
=
\operatorname{Re}
\left[
\frac{1-e^{-\mathrm{i}\Omega_j}}
{1+\beta e^{-\mathrm{i}\Omega_j}}
\right],
\qquad
\mathrm{SNR}_j
=
\frac{|d_j|\mu_j\sqrt{N_d/2}}
{\sigma_{\tilde J}},
\end{equation}
where $\mu_j$ represents the in-phase response of the high-pass filter defined in Eq.~\eqref{eq:esc_filter} at the perturbation frequency $\Omega_j$.

Assuming that the cost varies approximately linearly with the gain over the perturbation range, the periodic gain perturbation produces an approximately sinusoidal variation in the measured cost. Over one complete dither period of $N_d$ segments, the perturbation-correlated signal accumulates proportionally to $N_d/2$, whereas the standard deviation of uncorrelated background fluctuations increases proportionally to $\sqrt{N_d/2}$. Their ratio gives the SNR scaling in Eq.~\eqref{eq:esc_snr}. This approximation assumes that the filtered background fluctuations are sufficiently weakly correlated between segments.

A candidate gain is selected for adaptation when
\begin{equation}
\label{eq:esc_selection}
\mathrm{SNR}_j
\geq
\mathrm{SNR}_{\min},
\end{equation}
where $\mathrm{SNR}_{\min}>0$ is a prescribed selection threshold. A lower threshold allows more gains to be adapted but increases the risk of unreliable gradient estimates, whereas a higher threshold provides more conservative selection at the expense of potentially excluding useful gains.

In this study, $\mathrm{SNR}_{\min}=3$ is adopted as a conservative engineering criterion, motivated by the conventional three-standard-deviation rule. If the demodulated background fluctuations are approximately Gaussian and their standard deviation is accurately represented by the SNR approximation, this threshold corresponds to a nominal one-sided probability of approximately $0.13\%$ of estimating an incorrect gradient direction. This probability is conditional on the assumed noise model and is not treated as a guaranteed error rate.

The set of gains selected for adaptation is therefore defined as
\begin{equation}
\label{eq:esc_adapted_set}
\mathcal{A}
=
\left\{
j:\mathrm{SNR}_j\geq\mathrm{SNR}_{\min}
\right\}.
\end{equation}
Only gains indexed by $\mathcal{A}$ are updated during ESC training, while the remaining gains are held at their initial values.

For each selected gain, the adaptation gain $\kappa_j$ is subsequently determined to limit the magnitude of gain adjustments, subject to two constraints: (i) the predicted movement of the normalized nominal gain over one dither period is limited to $\Delta_c$; and (ii) the magnitude of an individual update caused by the observed background fluctuations is limited to $\Delta_{\mathrm{bg}}$.

Under this linearity assumption and the standard ESC averaging approximation, the periodic gain perturbation produces a cost variation with amplitude approximately equal to $|d_j|$, as estimated from the paired perturbations in Eq.~\eqref{eq:calibration_signal}. Correlating the filtered cost variation with the demodulation signal $a_j\cos(\Omega_j k)$ gives the predicted magnitude of the normalized nominal gain movement over one dither period,
\begin{equation}
\label{eq:esc_nominal_motion}
\Delta c_j^{\mathrm{pred}}
\approx
\frac{N_d}{2}
\kappa_j a_j |d_j|\mu_j.
\end{equation}
Imposing the prescribed limits on the predicted gain movement and the background-induced update gives
\begin{equation}
\label{eq:esc_rate}
\kappa_j
=
\min
\left\{
\frac{2\Delta_c}
{N_d a_j |d_j|\mu_j},
\;
\frac{\Delta_{\mathrm{bg}}}
{a_j\max_k|\tilde{J}_k^{\mathrm{bg}}|}
\right\}.
\end{equation}
When different performance objectives or objective weightings are prescribed, the calibration is repeated for the corresponding performance measurements, since both the gain-induced variations and the background fluctuations depend on the selected objective.

%%%
\subsection{Multi-Objective Cost Function}
\label{sec:cost}

The segment-wise ESC procedure developed in Section~\ref{subsec:esc_tuning} adjusts controller gains to minimize a prescribed scalar performance cost. For a given performance objective, the cost is denoted by $J(\bm{\theta}_{\mathrm{all}})$, where $\bm{\theta}_{\mathrm{all}}$ is the controller-gain vector defined in Eq.~\eqref{eq:theta_all}. During ESC tuning, this cost is evaluated through the segment performance measurements $J_k$ over the scoring window defined in Eq.~\eqref{eq:window_0}. For single-objective tuning of the speed controller, the mean-square speed-tracking error is adopted,
\begin{equation}
\label{eq:Jspeed}
J_k^u
=
\frac{1}{T_s}
\int_{\mathcal{W}_k}
e_u^2(t)\,dt,
\end{equation}
where $e_u(t)$ is the speed-tracking error defined in Section~\ref{subsec:pid_control}.

In many engineering applications, controller performance involves multiple objectives that may conflict with one another \citep{marler2010weighted,Wang2018}. In the present study, two competing objectives are considered: tracking performance and turning-induced heel. The tracking objective accounts for speed and heading tracking errors, whereas the heel objective considers the LF roll response associated with vessel manoeuvring. The latter is selected to characterize turning-induced heel rather than wave-frequency roll.

To account for differences in the physical units and characteristic magnitudes of the performance variables, a bounded penalty function is introduced as
\begin{equation}
\label{eq:hill}
h(\xi;M)
=
\frac{\xi^2}{\xi^2+M^2},
\qquad
0\leq h(\xi;M)<1,
\end{equation}
where $M>0$ is the characteristic scale of the corresponding performance variable. Using the speed and heading tracking errors defined in Section~\ref{subsec:pid_control}, the instantaneous tracking penalty is formulated as
\begin{equation}
\label{eq:Jerror_inst}
L_{\mathrm{tracking}}(t)
=
\frac{1}{2}
\left[
h(e_u(t);M_{\mathrm{spd}})
+
h(e_\psi(t);M_{\mathrm{head}})
\right],
\end{equation}
where $M_{\mathrm{spd}}$ and $M_{\mathrm{head}}$ are the characteristic scales for speed and heading errors, respectively, which specify the error magnitudes at which the corresponding penalties reach $0.5$, thereby balancing their contributions to the tracking objective. The corresponding segment tracking cost is obtained by averaging the instantaneous tracking penalty over the scoring window $\mathcal{W}_k$,
\begin{equation}
\label{eq:Jtracking}
J_{\mathrm{tracking},k}
=
\frac{1}{T_s}
\int_{\mathcal{W}_k}
L_{\mathrm{tracking}}(t)\,dt.
\end{equation}
The turning-induced heel penalty is defined using the LF roll angle $\phi_{LF}(t)$ and the bounded penalty function in Eq.~\eqref{eq:hill},
\begin{equation}
\label{eq:Jheel_inst}
L_{\mathrm{heel}}(t)
=
h\bigl(\phi_{LF}(t);\phi_{\lim}\bigr),
\end{equation}
where $\phi_{\lim}$ is the characteristic heel scale. The corresponding segment cost is
\begin{equation}
\label{eq:Jheel}
J_{\mathrm{heel},k}
=
\frac{1}{T_s}
\int_{\mathcal{W}k}
L_{\mathrm{heel}}(t)\,dt.
\end{equation}

Since the tracking and turning-induced heel objectives have different characteristic magnitudes, directly combining their costs may cause the objective with the larger numerical scale to dominate the optimization. To mitigate this scale dependence and enable a more meaningful weighting of the two objectives, their time-averaged costs are normalized by the corresponding reference costs evaluated at the initial controller gains. Specifically,
\begin{equation}
\label{eq:Jnormalized}
\hat J_i(\bm{\theta}_{\mathrm{all}})
=
\frac{J_i(\bm{\theta}_{\mathrm{all}})}
{J_i^{\mathrm{ini}}},
\qquad
i\in\{\mathrm{tracking},\mathrm{heel}\},
\end{equation}
where $J_i(\bm{\theta}_{\mathrm{all}})$ denotes the time-averaged cost of objective $i$, and $J_i^{\mathrm{ini}}>0$ is its reference value obtained using the initial controller gains. This normalization expresses both objectives relative to their initial performance, allowing their changes to be compared on a common dimensionless scale during ESC tuning.

The resulting normalized objectives are combined into a weighted cost function for ESC tuning ~\citep{marler2010weighted}:
\begin{equation}
\label{eq:Jtotal}
J(\bm{\theta}_{\mathrm{all}};w)
=
w\hat J_{\mathrm{tracking}}(\bm{\theta}_{\mathrm{all}})
+
(1-w)\hat J_{\mathrm{heel}}(\bm{\theta}_{\mathrm{all}}),
\qquad w\in[0,1],
\end{equation}
where $w$ determines the relative weighting between tracking performance and turning-induced heel.

For each prescribed weighting parameter $w$, the ESC procedure produces a candidate controller-gain vector $\bm{\theta}^{\mathrm{ESC}}(w)$. Varying $w$ yields different gain configurations, allowing the trade-off between tracking performance and turning-induced heel to be investigated.

The resulting controllers are compared in terms of their performance in the two objectives. A controller is considered Pareto dominated if another candidate achieves no worse performance in either objective and strictly better performance in at least one. The remaining nondominated candidates form an empirical Pareto set, representing the available trade-offs between tracking performance and turning-induced heel \citep{marler2010weighted}.

Selecting a controller from these trade-offs depends on the relative importance assigned to the two objectives. When this preference is specified, the controller with the lowest weighted cost can be selected. However, when the preference is uncertain, a compromise controller can be identified using the minimax-regret criterion \citep{savage1951theory}.

Let $\mathcal{P}\subseteq[0,1]$ denote the prescribed set of weighting values, and let $\mathcal{C}=\{\bm{\theta}^{\mathrm{ESC}}(w):w\in\mathcal{P}\}$ denote the corresponding set of candidate controllers. For each candidate $\bm{\theta}\in\mathcal{C}$, the regret under weighting $w\in\mathcal{P}$ is defined as
\begin{equation}
\label{eq:regret}
R(\bm{\theta},w)
=
J(\bm{\theta};w)
-
\min_{\bm{\theta}'\in\mathcal{C}}
J(\bm{\theta}';w),
\end{equation}
where $J(\bm{\theta};w)$ denotes the normalized multi-objective cost evaluated under common test conditions. The regret measures the additional cost incurred by selecting $\bm{\theta}$ instead of the best available candidate for that weighting.

The compromise controller is then selected by minimizing the maximum regret across all prescribed weightings,
\begin{equation}
\label{eq:minimax_regret}
\bm{\theta}^{\mathrm{comp}}
=
\arg\min_{\bm{\theta}\in\mathcal{C}}
\max_{w\in\mathcal{P}}
R(\bm{\theta},w).
\end{equation}
This criterion selects the candidate controller with the smallest worst-case regret across the prescribed weightings, providing a systematic compromise when the relative importance of the objectives is uncertain.

\section{Numerical Experiments}
\label{sec:simulation}

This section evaluates the proposed offline controller-tuning framework through two numerical experiments. The first investigates the influence of the tuning environment by comparing controller gains obtained in calm water and stochastic waves, with particular attention to tracking performance and propulsion response. The second examines multi-objective controller tuning under stochastic wave excitation, focusing on the trade-off between tracking performance and turning-induced heel during large course changes. In both experiments, the tuned controllers are evaluated on independent wave realizations to assess their performance under conditions not used for tuning.

\subsection{Simulation Setup}
\label{subsec:sim_setup}

The numerical experiments are based on the polar research vessel \textit{Akademik Tryoshnikov}, which served as the platform for the Antarctic Circumnavigation Expedition (ACE)~\citep{Schmaleetal2019,derkanietal2021}. Its principal particulars and hull geometry are obtained from publicly available full-scale data. Since complete design and inertial properties are unavailable, the inertia and stiffness parameters used in the wave-response model are adopted from the Arc7 ice-class hull model investigated by \citet{yu2026nemoh}. The vessel is represented as an underactuated surface vessel equipped with a diesel-electric propulsion system, with its propulsion characteristics represented by a calibrated surrogate model.

\begin{figure}[pos=htbp]
  \centering
  \includegraphics{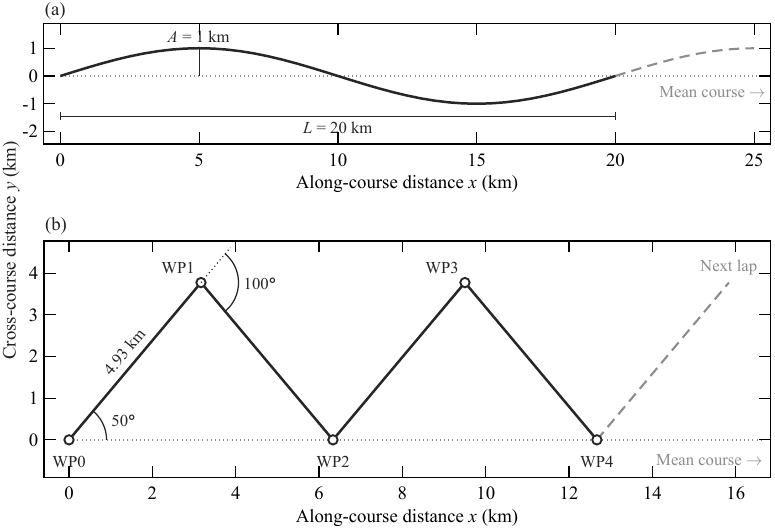}
  \caption{Reference paths. (a) Reference Path 1, a sinusoidal path; dashed: start of the next wavelength. (b) Reference Path 2, a zig-zag route derived from the IEC~62065 track-control test track. One lap (WP0-WP4) consists of four legs of equal length at alternating course deviations of $\pm50^\circ$ about the mean course, so that each course change is $100^\circ$, and corresponds to one evaluation segment. The lap is repeated along the mean course; dashed: first leg of the next lap.}
  \label{fig:route}
\end{figure}

Two reference paths are considered to investigate single-objective and multi-objective controller tuning, consistent with the reference-signal averaging considerations introduced in Eq.~\eqref{eq:reference_average}. Reference Path 1 is a sinusoidal trajectory adopted to compare single-objective ESC tuning in calm water and stochastic waves (Section~\ref{subsec:performance_comparison}). The path is defined by $y=A\sin(2\pi x/L)$ with $A=1000$~m and $L=20000$~m, sampled as a polyline with waypoints spaced 500~m apart along $x$ (Fig.~\ref{fig:route}a).
Sinusoidal paths have been used in marine vessel tracking studies to evaluate control performance under varying course demands and environmental disturbances \citep{Belleter2019,ZhangLiuAnderlini2022}. The smooth and repeated course variations facilitate comparable performance evaluations across successive ESC segments without introducing abrupt heading changes.

Reference Path 2 is a periodic zig-zag trajectory derived from IEC~62065, the track-control standard of the International Electrotechnical Commission~\citep{IEC62065}, adopted for multi-objective ESC tuning to investigate the trade-off between tracking performance and turning-induced heel during large course changes (Section~\ref{subsec:tradeoff}). Its repeated manoeuvring pattern facilitates segment-wise performance evaluation, while the large course changes introduce the steering demands needed to examine the competing objectives, as illustrated in Fig.~\ref{fig:route}b.

Both reference paths use the same LOS guidance parameters, with a look-ahead distance of $D_{ah}=1,250$~m, a waypoint acceptance radius of $R_{\mathrm{acc}}=250$~m, and a desired vessel speed of $U_d=7$~m/s, or about 13.6~knots. The maximum open-water speed of the \textit{Akademik Tryoshnikov} is 16~knots, which corresponds to the shaft-speed limit of 150~rpm in the model. The desired speed is therefore set below this maximum, so that the shaft speed retains a margin for speed regulation in waves. In the comparison of tuning environments, additional sensor noise is excluded to isolate the influence of wave-induced feedback fluctuations on controller tuning. In the multi-objective experiment, the measurement noise specified in Section~\ref{subsec:integrated_model} is included to investigate the combined effects of wave-induced feedback fluctuations and measurement uncertainty on controller performance.

\subsubsection{Vessel Specifications and Stochastic Wave Conditions}

The vessel parameters, stochastic wave conditions, and numerical solver settings used in the simulations are summarized in Table~\ref{tab:sim_params}.

\begin{table}[htb!]
\centering
\caption{Simulation parameters and environmental settings.}
\label{tab:sim_params}
\begin{tabular}{lcc}
\hline
\textbf{Parameter} & \textbf{Symbol} & \textbf{Value} \\ \hline
\multicolumn{3}{c}{\textit{Vessel Specifications (Akademik Tryoshnikov)}} \\
Length overall & $L_{OA}$ & 133.53 m \\
Molded breadth & $B$ & 23.25 m \\
Design draught & $T_d$ & 8.50 m \\
Full-scale displacement & $\Delta_d$ & 16,539 t \\
Model displaced mass & $\Delta$ & 11,515 t \\
\multicolumn{3}{c}{\textit{Environmental Conditions (JONSWAP)}} \\
Significant wave height & $H_s$ & 2.5 m \\
Peak wave period & $T_p$ & 8.0 s \\
Peak enhancement factor & $\gamma$ & 3.3 \\
Frequency range & $\omega$ & $[0.049,\,\pi]$ rad/s \\
Number of components / spacing & $N_\omega$, $\Delta\omega$ & 64, 0.0491 rad/s \\ \hline
\multicolumn{3}{c}{\textit{Solver Settings}} \\
Time step & $\Delta t$ & 0.05 s (logged at 0.1 s) \\
Solver type & -- & Fixed-step Runge--Kutta (4th order) \\
\hline
\end{tabular}
\end{table}

Independent stochastic wave realizations are generated by varying the random phases of the discretized JONSWAP spectral components, as described in Section~\ref{subsubsec:wave_reconstruction}. The frequency spacing gives a nominal wave recurrence time of $T_w\approx128$~s, which is used to determine the ESC scoring-window duration in Section~\ref{subsec:esc_tuning}, without implying periodic repetition of the closed-loop vessel response. For both numerical experiments, the global wave-propagation direction is fixed at $0^\circ$, aligned with the mean course of the reference paths.

\subsubsection{Common ESC Parameters and Evaluation Protocol}
\label{subsubsec:esc_params}

Both numerical experiments employ the segment-wise ESC procedure developed in Section~\ref{subsec:esc_tuning}. The initial controller gains are set to $(K_{p,u},K_{i,u},K_{d,u},K_{p,\psi})=(20,\,0.25,\,0.995,\,2)$, obtained through preliminary manual tuning. Each evaluation segment has a duration of $T_{\mathrm{seg}}=3{,}000$~s, with performance evaluated over the final $T_s=18T_w\approx2{,}300$~s to reduce the influence of transient responses following gain updates.

The ESC parameters are set to $N_d=32$ segments per dither period and $\beta=0.5$ for the high-pass filter. The SNR-based calibration employs $N_b=6$ fixed-gain segments to estimate background fluctuations. The adaptation limits in Eq.~\eqref{eq:esc_rate} are prescribed as $\Delta_c=0.35$ and $\Delta_{\mathrm{bg}}=0.032$. Following an initial evaluation segment, the controller gains are updated $N_u=32$ times. The candidate gains, their admissible intervals, and perturbation parameters are specified separately for each numerical experiment in Sections~\ref{subsec:performance_comparison} and~\ref{subsec:tradeoff}.

To assess the performance of the tuned controllers, the final gains are held fixed and evaluated on five independent wave realizations that are not used during calibration or ESC training. Each test consists of a 700~s warm-up followed by ten consecutive segments of 3,000~s. All controllers are tested using the same wave realizations and initial vessel states, allowing paired comparisons of their closed-loop performance.

\subsection{Single-Objective ESC Tuning in Calm Water and Stochastic Waves}
\label{subsec:performance_comparison}

This experiment employs Reference Path 1 to compare single-objective ESC tuning in calm water and stochastic waves. The speed-controller gains are tuned in both environments using the same initial gains, ESC parameters, and number of updates, with the speed-tracking cost in Eq.~\eqref{eq:Jspeed} as the performance cost. Both procedures use one prescribed training realization, with wave excitation omitted in the calm-water case. During training, the calm-water simulations restart from the same initial vessel state for each evaluation segment, whereas wave-based tuning proceeds continuously without resetting the vessel state or wave excitation. The resulting controllers, together with the initial hand-tuned controller, are then evaluated under identical independent wave conditions.

The speed proportional and integral gains, $K_{p,u}$ and $K_{i,u}$, are considered as candidate parameters, with admissible intervals $[2.5,37.5]$ and $[0.025,0.75]$, dither amplitudes $s_ja_j$ of 2.8 and 0.029, and perturbation cycle numbers $b_j=13$ and 11 over $N_d=32$ segments, respectively. The SNR-based calibration under stochastic waves yields $\mathrm{SNR}=15.2$ for $K_{p,u}$ and $0.13$ for $K_{i,u}$, so only $K_{p,u}$ exceeds the threshold $\mathrm{SNR}_{\min}=3$ and is selected for adaptation, while $K_{i,u}=0.25$, $K_{d,u}=0.995$, and $K_{p,\psi}=2$ remain fixed in both tuning procedures. To ensure comparable adaptation settings, the larger value of $|d_j|$ obtained from the calm-water and wave-based calibrations is used in Eq.~\eqref{eq:esc_rate} to determine a common adaptation gain.

Table~\ref{tab:2env_comp_table} summarizes the resulting controller gains and their performance on the independent test realizations. The speed proportional gain decreases from its initial value of $20$ to $18.29$ after calm-water tuning and to $6.06$ after wave-based tuning. The wave-tuned controller achieves the lowest speed-tracking cost across all five independent test realizations. Relative to the wave-tuned controller, the mean speed-tracking cost is 2.6\% higher for the calm-water-tuned controller and 3.1\% higher for the initial controller. The corresponding speed RMSE values are 0.170~km/h and 0.168~km/h for the calm-water-tuned and wave-tuned controllers, respectively, indicating a relatively small improvement in tracking accuracy.

\begin{table}[htbp]
\centering
\caption{Frozen-gain performance of the initial, calm-water-tuned, and wave-tuned controllers on independent wave realizations. The rate-limited fraction refers to the first test segment.}

\label{tab:2env_comp_table}
\small
\setlength{\tabcolsep}{4pt}
\begin{tabular*}{\linewidth}{@{\extracolsep{\fill}}lcccccc}
\hline
Controller
& $K_{p,u}$
& $K_{i,u}$
& $J^u$
& $\sigma_n$
& RMS$(\dot n)$
& Rate-limited \\
& [rpm\,s/m]
& [rpm/m]
& [$(\mathrm{m/s})^2$]
& [rpm]
& [rpm/s]
& [\%] \\
\hline
Initial PID
& 20.00 & 0.250 & 0.002247 & 0.898 & 0.815 & 39.4 \\
Calm-water ESC PID
& 18.29 & 0.250 & 0.002236 & 0.831 & 0.780 & 35.7 \\
Wave-tuned ESC PID
& 6.06 & 0.250 & \textbf{0.002180}
& \textbf{0.283} & \textbf{0.466} & \textbf{11.2} \\
\hline
\end{tabular*}
\end{table}

More substantial differences are observed in the propeller response. Compared with the wave-tuned controller, the calm-water-tuned controller exhibits approximately 2.9 times greater shaft-speed variability and 1.7 times greater root-mean-square shaft-speed rate, RMS$(\dot n)$. The proportion of time spent at the shaft-speed rate limit also increases from approximately 11\% to 36\%. Meanwhile, the mean vessel speed, shaft power, and energy consumption differ by less than 0.01\% among the three controllers.

\subsection{Multi-Objective ESC Tuning: Tracking--Heel Trade-off}
\label{subsec:tradeoff}

This experiment employs Reference Path 2, whose large course changes may require more aggressive steering to improve tracking, potentially increasing turning-induced heel through yaw-roll coupling~\citep{vanAmerongen1990,OhSun2010,ZhangSunLiu2017}. The route is derived from Scenario~3 of the IEC~62065 track-control standard, which specifies alternating course directions of $040^\circ$ and $140^\circ$, corresponding to $100^\circ$ course changes \citep{IEC62065,Zhang2021}. It is extended into a periodic zig-zag pattern consisting of four legs per lap, each of length 4926.8~m and alternating by $\pm50^\circ$ about the mean course, as illustrated in Fig.~\ref{fig:route}. The leg length is selected such that, with the initial controller gains and desired speed $U_d=7$~m/s, one lap takes approximately $T_{\mathrm{seg}}=3000$~s, providing approximately repeated manoeuvring demands across successive ESC evaluation segments.

The multi-objective cost defined in Eq.~\eqref{eq:Jtotal} is adopted to balance tracking performance and turning-induced heel, with the tracking and heel costs defined in Eqs.~\eqref{eq:Jtracking} and~\eqref{eq:Jheel}. The characteristic scales are set to $M_{\mathrm{spd}}=0.5~\mathrm{m/s}$ and $M_{\mathrm{head}}=15^\circ$, which are selected as representative error scales for the simulated manoeuvring conditions rather than strict tracking limits, and $\phi_{\lim}=10^\circ$, motivated by the turning-heel criterion for passenger ships in the 2008 IS Code~\citep{IMO2008}. The heel scale defines the characteristic scale of the penalty rather than imposing a hard stability constraint. Angular quantities are expressed in radians in the numerical implementation.

For heel angles small relative to $\phi_{\lim}$, the penalty satisfies
\begin{equation}
h(\phi_{LF};\phi_{\lim})
\approx
\frac{\phi_{LF}^2}{\phi_{\lim}^2}.
\end{equation}
Since the simulated turning-induced heel remains substantially below the prescribed characteristic scale, the heel penalty operates predominantly in this approximately quadratic regime. Furthermore, normalizing the heel cost by its reference value at the initial controller gains, as defined in Eq.~\eqref{eq:Jnormalized}, reduces the influence of the absolute penalty scale on the weighted objective.

The reference costs evaluated at the initial controller gains are $J_{\mathrm{tracking}}^{\mathrm{ini}}=0.1673$ and $J_{\mathrm{heel}}^{\mathrm{ini}}=0.00276$. These values are held fixed across all weighting parameters and used to normalize the respective objectives according to Eq.~\eqref{eq:Jnormalized}, both during ESC tuning and independent performance evaluation.

Nine weighting values, $w=0,\tfrac18,\tfrac28,\ldots,1$, are considered, ranging from heel-only optimization ($w=0$) to tracking-only optimization ($w=1$). For each weight, ESC tuning is performed independently over $N_r=5$ training wave realizations.

All four controller gains, $K_{p,u}$, $K_{i,u}$, $K_{d,u}$, and $K_{p,\psi}$, are initially considered for adaptation, with admissible intervals $[2.5,37.5]$, $[0.025,0.75]$, $[0.2,6]$, and $[0.5,4.5]$, dither amplitudes $s_ja_j$ of $2.8$, $0.087$, $0.1$, and $0.32$, and perturbation cycle numbers $b_j=13$, $11$, $9$, and $15$, respectively. The SNR-based calibration, performed separately for each weight, shows that $K_{p,u}$, $K_{i,u}$, and $K_{p,\psi}$ satisfy the selection criterion $\mathrm{SNR}_j\geq3$ for all nine weights, whereas $K_{d,u}$ remains below the threshold ($\mathrm{SNR}\leq0.002$) and is held at its initial value. For each weight, the final gains obtained from the five training realizations are averaged to produce one candidate controller, $\bm{\theta}^{\mathrm{ESC}}(w)$, and the nine candidates are then evaluated on the independent test realizations.

\begin{figure*}[pos=htb!]
\centering
\includegraphics[width=\textwidth]{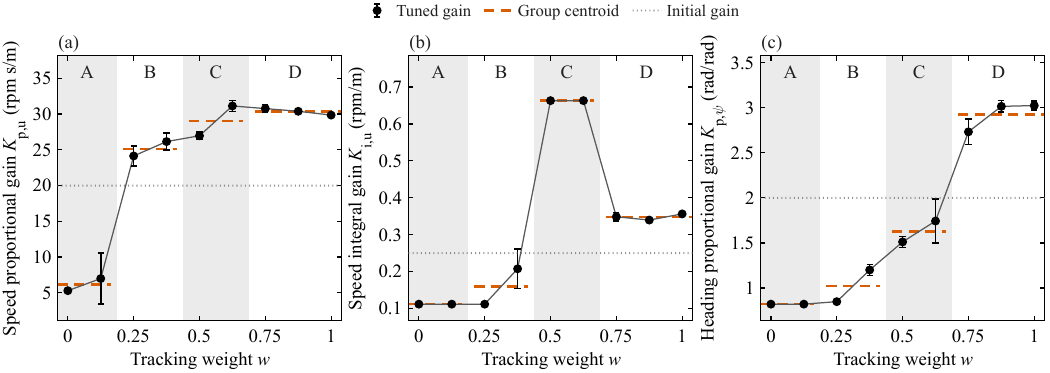}
\caption{Groups of optimized controller gains for different tracking weights. Centroids represent the mean gains within each group, and distance reductions are measured relative to the initial controller gains in normalized gain space.}
\label{fig:gains}
\end{figure*}

The influence of the weighting parameter on the optimized controller gains is illustrated in Fig.~\ref{fig:gains}. As the relative importance of tracking performance increases, the heading proportional gain generally increases, indicating more aggressive steering. Similar gain configurations form four groups, summarized in Table~\ref{tab:gain_groups}.

\begin{table}[htb!]\rmfamily
\centering
\caption{\rmfamily Gain groups of the tuned gains. Centroid: mean tuned gains of the group, proposed as the training start for weights within the group. Last column: reduction of the normalized distance from the start to the tuned gains of the group members when the centroid replaces the initial gains as the start.}
\label{tab:gain_groups}
\begin{tabular}{lccccc}
\hline
Group & $w$ & $K_{p,u}$ [rpm\,s/m] & $K_{i,u}$ [rpm/m] & $K_{p,\psi}$ [rad/rad]& Distance reduction [\%] \\
\hline
A & 0, 0.125 & 6.15 & 0.112 & 0.82 & 96 \\
B & 0.25, 0.375 & 25.2 & 0.160 & 1.02 & 72--79 \\
C & 0.5, 0.625 & 29.1 & 0.663 & 1.63 & 88--89 \\
D & 0.75, 0.875, 1 & 30.3 & 0.348 & 2.93 & 84--93 \\
\hline
\end{tabular}
\end{table}

In the normalized gain space, the distance from each group centroid to the tuned gains of its members is 72--96\% shorter than that from the initial gains (Table~\ref{tab:gain_groups}). Since the ESC step size is fixed by the calibration, a start closer to the optimum needs fewer updates, so the group centroids can serve as starting points for further training when a new preference falls within the weight range of a group. The frozen tests also show that gains within a group are interchangeable: the gains of any member, used at any weight of the same group, increase the cost by at most 0.54\%. Two caveats apply. Some adapted gains stop at the limits imposed by the projection in Eq.~\eqref{eq:esc_projection}: all three gains at $w=0$, $K_{i,u}$ and $K_{p,\psi}$ at $w=0.125$, and $K_{i,u}$ in groups B (lower limit) and C (upper limit). The results at these weights and some group boundaries are therefore set by the admissible intervals. In addition, the centroid of group D is computed from gains that are still changing at the end of training.

\begin{figure}[pos=htb!]
\centering
\includegraphics{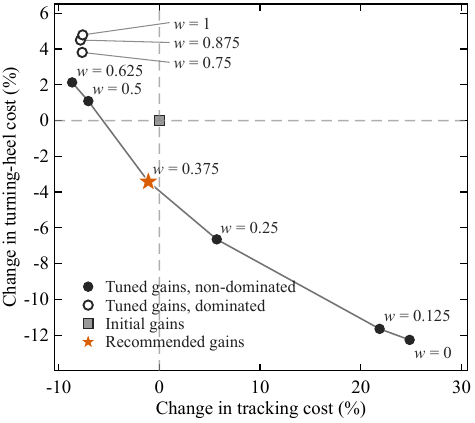}
\caption{Empirical trade-off between tracking performance and turning-induced heel for the ESC-tuned controllers evaluated on five independent wave realizations. Changes in both costs are expressed relative to the initial controller. Filled markers indicate nondominated controllers, open markers indicate dominated controllers, and the star identifies the compromise controller obtained at $w=0.375$ using the minimax-regret criterion.}
\label{fig:pareto}
\end{figure}

The frozen tests yield an empirical tracking-heel trade-off curve (Fig.~\ref{fig:pareto}). The gains obtained for $w=0.75$, 0.875, and 1 are dominated by those obtained for $w=0.625$ and are therefore excluded from the nondominated subset. Among the nondominated gains, as $w$ increases from 0 to 0.625, the tracking cost relative to the initial gains decreases from $+24.8\%$ to $-8.7\%$ and the heel cost increases from $-12.3\%$ to $+2.1\%$, consistently in all five test realizations. The maximum turning-induced heel is reduced by 24\% at $w=0$ and increased by 6\% at $w=1$. At the heel-oriented end, the standard deviation of the shaft speed is also reduced by 46\% and the propulsion energy by 3.8\%, at the cost of a 34\% larger speed RMSE.

The initial controller is not Pareto optimal among the evaluated configurations: the controller obtained at $w=0.375$ reduces the tracking and turning-induced heel costs by $1.1\%$ and $3.4\%$, respectively, relative to the initial gains. Applying the minimax-regret criterion of Eq.~\eqref{eq:minimax_regret} to the frozen-gain evaluations across all nine weights, this controller achieves the smallest worst-case regret of $0.091$, compared with $0.126$ for the initial controller and $0.137$ for the next-best tuned controller ($w=0.5$). Moreover, it achieves a lower weighted cost than the initial controller across all prescribed weights, with improvements ranging from $1.1\%$ to $3.4\%$. It is therefore selected as the compromise controller among the evaluated candidates.

\section{Discussion and Limitations}
\label{subsec:discussion}

The numerical results demonstrate the potential of the proposed offline controller-tuning framework under stochastic wave excitation. Previous studies have established the potential of ESC for model-free controller tuning \citep{killingsworth2006pid}, while stochastic disturbances can complicate gradient estimation and adaptation \citep{stankovic2010extremum}. In the first experiment, wave-based tuning produced a substantially smoother propeller response than calm-water tuning, while maintaining comparable speed-tracking and energy performance. The lower speed proportional gain obtained through wave-based tuning is consistent with reduced sensitivity to wave-frequency fluctuations in the measured feedback signals. Although the improvement in speed-tracking accuracy was relatively small, the reductions in shaft-speed fluctuations and rate-limited operation suggest that accounting for wave-induced feedback variations during tuning can mitigate unnecessary actuator activity. These findings highlight the importance of considering the operating environment when evaluating controller performance, consistent with the broader challenges of marine motion control under environmental disturbances \citep{fossen2011handbook}.

The second experiment further demonstrates the applicability of the proposed framework to controller tuning with competing performance requirements. The conflict between tracking performance and turning-induced heel is associated with manoeuvring-induced yaw-roll coupling, which has motivated previous studies on roll-constrained path following and roll stabilization \citep{perez2005ship,ZhangSunLiu2017,Liang2023}. In the present study, varying the relative weighting of these objectives produced different controller configurations and an empirical tracking-heel trade-off, consistent with the weighted-sum approach to multi-objective optimization \citep{marler2010weighted}. The SNR-based calibration also excluded the speed derivative gain, whose effect could not be sufficiently distinguished from background fluctuations, thereby avoiding unreliable updates of that gain. The initial hand-tuned controller was not Pareto optimal: the compromise controller identified by the minimax-regret criterion~\citep{savage1951theory} improved both objectives across all prescribed weights, rather than sacrificing one to improve the other. This suggests that the proposed framework can support controller selection when the relative importance of competing objectives is uncertain.

Several limitations should be acknowledged. The numerical experiments employ a two-time-scale vessel-wave model in which first-order wave-induced motions influence the closed-loop system only through the feedback measurements, and the HF responses are represented using linear potential-flow theory~\citep{fossen2011handbook,skejic2008unified,perez2005ship}. Nonlinear wave-vessel interactions, second-order wave-drift forces, wind, and current disturbances are excluded, and the evaluations are restricted to a single sea state with unidirectional (long-crested) waves and a fixed propagation direction. Consequently, the observed performance improvements cannot be directly generalized to other vessels or environmental conditions without further investigation.

The proposed ESC procedure also relies on assumptions concerning closed-loop stability, performance averaging, and the detectability of gain-induced cost variations, which are closely related to the averaging and stability properties of ESC investigated in previous studies \citep{stankovic2010extremum}. The SNR-based calibration estimates gain sensitivity and background fluctuations under prescribed operating conditions, and the resulting adaptation gains remain fixed throughout training. Changes in local cost sensitivity may therefore affect the convergence behaviour. In particular, the controllers obtained for $w\geq0.75$ are Pareto dominated, while their gains were still changing at the end of the 32 updates, suggesting that incomplete convergence contributes to their dominance. Additional updates or adaptive step-size selection may therefore be necessary, although their effectiveness remains to be investigated.

The framework has also been evaluated only through numerical simulations rather than physical vessel experiments, and the comparison between calm-water and wave-based tuning employs different state-reset procedures, so the observed performance differences cannot be attributed exclusively to wave excitation. Further investigations involving additional sea states, wave directions, alternative tuning strategies, and experimental validation are required to establish the robustness and practical applicability of the proposed framework.

\section{Conclusion}
\label{sec:conclusion}

This paper developed an offline controller-tuning framework that integrates a two-time-scale vessel-wave simulation model with model-free ESC for PID gain optimization under stochastic wave excitation. The proposed segment-wise procedure enables gain adaptation during continuous simulated vessel operation without resetting the vessel state between successive evaluations. An SNR-based calibration procedure identifies gains suitable for adaptation and determines their adaptation gains by distinguishing gain-induced performance changes from stochastic background fluctuations.

Two numerical experiments were conducted to evaluate the proposed framework. In the single-objective experiment, tuning under stochastic waves reduced propeller shaft-speed variability and its root-mean-square rate by approximately 66\% and 40\%, respectively, compared with calm-water tuning, while reducing the speed-tracking cost by approximately 2.5\%. These results highlight the importance of accounting for wave-induced feedback fluctuations during controller tuning, particularly in reducing unnecessary actuator activity.

In the multi-objective experiment, a normalized performance cost was employed to balance tracking performance and turning-induced heel during large course changes. Varying the objective weighting revealed an empirical trade-off between the two competing objectives, with the tuned gains forming four groups within which they were interchangeable at a cost loss of at most 0.54\%. The controller obtained at $w=0.375$ improved both tracking and turning-induced heel performance relative to the initial controller and achieved the smallest worst-case regret among the evaluated candidates. These results demonstrate the potential of the proposed framework for controller-gain selection under competing performance requirements.

Within the numerical conditions investigated, the ability to adapt controller gains during continuous simulated vessel operation, together with SNR-based gain selection, provides a promising basis for online controller tuning. Whether these benefits persist across other sea states and wave directions, and whether gains tuned in simulation transfer to the real vessel, remain open. The fixed step size set by the calibration also limits convergence when the local cost sensitivity changes during training, indicating that the step size itself should adapt. The decisive test of the framework's practical value is therefore its performance during continuous operation of the real vessel.

\section*{Declaration of competing interest}
The authors declare that they have no known competing financial interests or personal relationships that could have appeared to
influence the work reported in this paper.

\section*{Data availability}
The simulation code and data supporting the findings of this study are available from the corresponding author upon reasonable request.

\section*{CRediT authorship contribution statement}
\textbf{Weijia Li:} Conceptualization, Methodology, Software, Investigation, Writing -- original draft.
\textbf{Ying Tan:} Supervision, Methodology, Writing -- review \& editing.
\textbf{Alessandro Toffoli:} Supervision, Writing -- review \& editing.

\section*{Acknowledgements}
The authors acknowledge support from the Australian Research Council (LP210200927, DP260101027).

\bibliographystyle{cas-model2-names}
\bibliography{references,References_ESC}

\end{document}